\documentclass[prl,twocolumn,showpacs,preprintnumbers,amsmath,amssymb,superscriptaddress]{revtex4}
\usepackage{amsmath,amssymb,bm,graphicx,epsfig,psfrag,ulem,color,natbib}

\newcommand{\bs}{\mathbf {s}}
\newcommand{\br}{\mathbf {r}}
\newcommand{\bv}{\mathbf {v}}
\newcommand{\etal}{{\it et al.}~}

\begin{document}

\title{Quasiclassical and ultraquantum decay of superfluid turbulence}

\author{A.~W.~Baggaley}\affiliation{School of Mathematics and Statistics, University of
Newcastle, Newcastle upon Tyne, NE1 7RU, UK}
\author{C.~F.~Barenghi}\affiliation{School of Mathematics and Statistics, University of
Newcastle, Newcastle upon Tyne, NE1 7RU, UK}
\author{Y.A.~Sergeev}\affiliation{School of Mechanical and Systems Engineering, Newcastle
University, Newcastle upon Tyne, NE1 7RU, UK}

\begin{abstract}
We address the question which, after a decade-long
discussion, still remains open: what is the nature of the
ultraquantum regime of decay of quantum turbulence? The model
developed in this work reproduces both the ultraquantum and the
quasiclassical decay regimes and explains their hydrodynamical
natures. In the case where turbulence is generated by forcing at
some intermediate lengthscale, e.g. by the beam of vortex rings in
the experiment of Walmsley and Golov [Phys. Rev. Lett. {\bf 100},
245301 (2008)], we explained the mechanisms of generation of both
ultraquantum and quasiclassical regimes. We also found that the
anisotropy of the beam is important for generating the large scale 
motion associated with the quasiclassical regime.
\end{abstract}


\maketitle
\newpage

The existence of a macroscopic complex order parameter in
superfluid helium ($^4$He and $^3$He) constrains the vorticity to
vortex lines, each line carrying one quantum of circulation
$\kappa$. This is in sharp contrast to ordinary fluids, where
vorticity is continuous. An important question is how quantum
turbulence compares to classical turbulence \cite{Vinen2002}.
Experiments in helium have revealed two regimes
\cite{Walmsley2008,Bradley2006,Bradley2011} of turbulent decay
characterized by $L \sim t^{-1}$ (ultraquantum) and $L \sim
t^{-3/2}$ (quasiclassical) behaviour, where $t$ is time and the
vortex line density (vortex length per unit volume) $L$ measures
the turbulence's intensity. In these two regimes the
kinetic energy (per unit mass) decays as
$E\sim t^{-1}$ and $E\sim t^{-2}$, respectively.
(Here it seems appropriate to point to the detailed
theoretical analysis~\cite{Adzhemyan1998} 
of energy decay in classical, viscous,
uniform and isotropic turbulence.) The second regime is thought to be
associated with the classical Kolmogorov distribution of kinetic
energy over the length scales, but the nature of the first regime
is still a mystery. Here we show that the first regime, associated
entirely with the Kelvin wave cascade along individual vortex
lines, takes place when the energy input at some intermediate
lengthscale is insufficient to induce the large-scale motion which
is associated with quasiclassical, ``Kolmogorov'' turbulence. In
other words, the first regime is a transient turbulent state which
decays before energy can be transferred to large scales by vortex
reconnections, which play a key role in this reverse energy
transfer.

Theoretical and experimental studies have revealed analogies
between superfluid turbulence and classical turbulence, notably
the same Kolmogorov energy spectrum in continually forced
turbulence \cite{Nore1997,Maurer1998,Kobayashi2005,Lvov2006,Salort2010}, 
as well as many dissimilarities and new effects.
Our concern is the decay of pure superfluid turbulence at
temperatures small enough that thermal excitations are negligible;
in the absence of viscous forces, in $^4$He the only mechanism
\cite{Vinen2002} to dissipate kinetic energy is phonon emission at
length scales much shorter than the average intervortex distance
$\ell \approx L^{-1/2}$. (In $^3$He-B, which is a fermionic
superfluid, the dissipation is thought to be associated with the
Caroli-Matricon mechanism~\cite{Kopnin1995} of energy loss from
short Kelvin waves into the quasiparticle bound states.) In this
limit turbulence reduces to a very simple form: a disordered
tangle of vortex lines, all of the same strength, moving in a
fluid without viscosity, but still retains the crucial features of
classical turbulence, the nonlinearities of the Euler equations
and the huge number of length scales which are excited.

By injecting negative ions in superfluid $^4$He in this
zero-temperature limit, Walmsley and Golov~\cite{Walmsley2008}
observed two regimes of turbulence decay corresponding to two
regimes of quantum turbulence discussed earlier in
Refs.~\cite{Volovik2003,Skrbek2006}. The negative ions (electron
bubbles) generated vortex rings \cite{Winiecki2000}; the rings
interacted with each other, forming a turbulent vortex tangle,
which, in the first regime, decayed as $L \sim t^{-1}$. The second
regime, characterized by $L \sim t^{-3/2}$, was observed if the
injection time was longer. The same $t^{-3/2}$ time dependence was
observed in the spin-down of a vortex lattice \cite{Walmsley2007},
and, at higher temperatures, during the decay of turbulence
initially generated by a towed grid \cite{Stalp1999}. Recently it
has been also modeled numerically \cite{Fujiyama2010}.

Walmsley and Golov argued that the second regime (which they
referred to as {\it Kolmogorov} or {\it quasiclassical} turbulence
\cite{Volovik2003,Skrbek2006}) is associated with the classical
Kolmogorov spectrum $E_k \sim k^{-5/3}$ at wavenumbers $k \ll
1/\ell$, whereas the first regime (called {\it Vinen} or {\it
ultraquantum} turbulence \cite{Volovik2003,Skrbek2006}) depends on
energy contained at smaller scales, $k \gg 1/\ell$. The ultraquantum
($L \sim t^{-1}$) and quasiclassical ($L \sim t^{-3/2}$) decay
regimes were also observed in $^3$He-B
by Bradley \etal~\cite{Bradley2006};
in this case the turbulence was generated by a vibrating grid
which sheds vortex loops in alternating directions.

Following Schwarz~\cite{Schwarz1988}, we have modeled vortex lines
as space curves $\bs=\bs(t,\xi)$ (where $\xi$ is arclength) which
move according to the classical Biot-Savart law
\begin{equation}
\frac{d{\bf s}}{dt}=-\frac{\kappa}{4 \pi} \oint_{\cal L} \frac{(\bs-\br) }
{\vert \bs - \br \vert^3}
\times {\bf d}\br,
\label{eq:BS}
\end{equation}
where the line integral extends to the entire vortex configuration
$\cal L$. Our model includes vortex reconnections and sound emission
(for details see
Refs~\cite{Baggaley-cascade,Baggaley-tree,Baggaley-long}).
The computational domain is a periodic box of size $D=0.03~\rm cm$.
Modeling the experiments~\cite{Walmsley2008}, the initial condition
represents the beam of vortex rings of radius
$R=6 \times 10^{-4}\,\rm cm$ injected up to time $t=0.1~\rm s$ with
initial velocity randomly confined within a $\pi/10$ angle.

The numerical techniques to de-singularize the Biot-Savart
integral, discretize the vortex filaments over a large variable
number of vortex points $\bs_j$, and perform vortex reconnections
are standard in the literature~\cite{Schwarz1988}. Details of our
algorithms are in our previous
papers~\cite{Baggaley-cascade,Baggaley-tree,Baggaley-long}, which
also describe the tree algorithm used to speed up the calculation
of Biot-Savart integrals. Our model includes small energy losses
at vortex reconnections, as described by more microscopic
calculations based on the Gross-Pitaevskii
equation~\cite{Leadbeater2001}. Energy losses due to sound
emission are modelled by the spatial
discretization~\cite{Baggaley-cascade}: vortex points are removed
if the local wavelength is smaller than a given minimum resolution
$\delta=5\times10^{-4}\,{\rm cm}$. In our calculations, the time
step is $\Delta t=5\times10^{-6}\,{\rm s}$. We have tested that 
the ultraquantum and quasiclassical behaviours remain the same if 
$\delta$ is halved.

\begin{figure}[h]
\begin{center}
\includegraphics[width=0.40\textwidth]{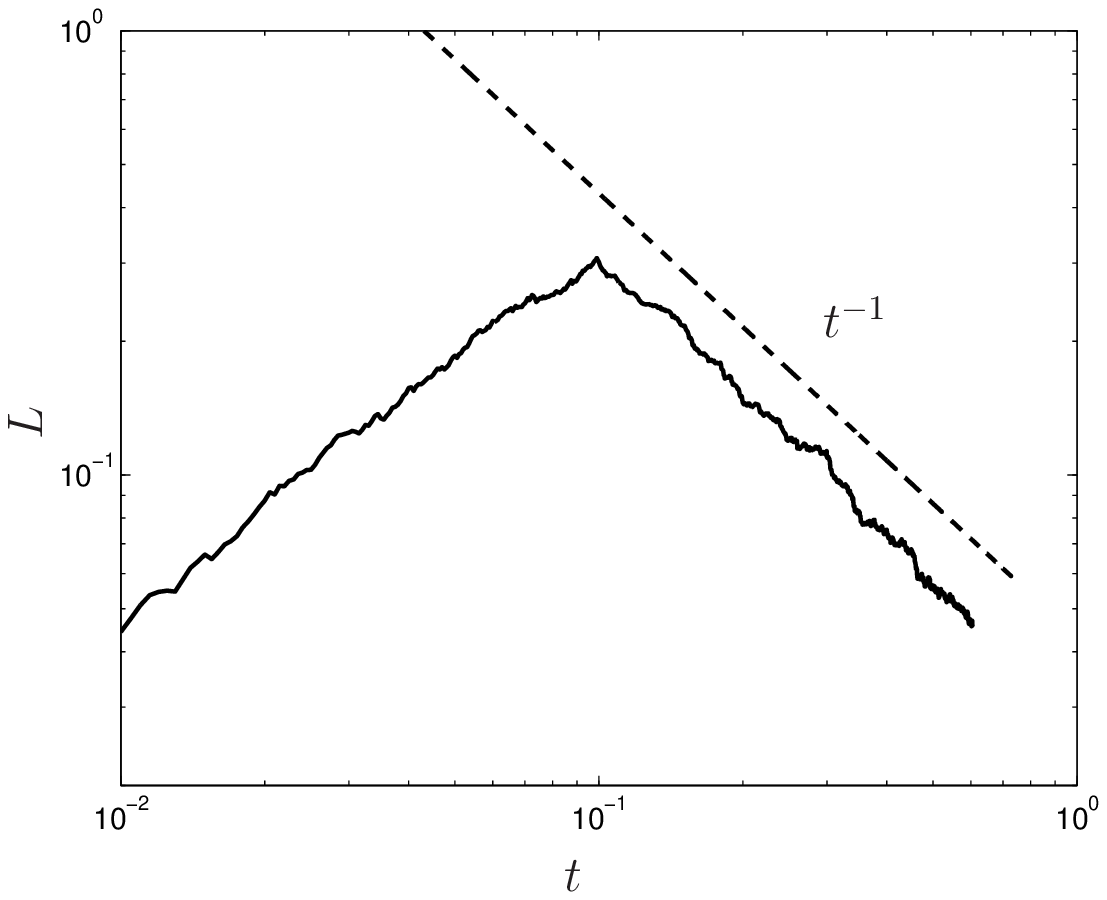}
\includegraphics[width=0.40\textwidth]{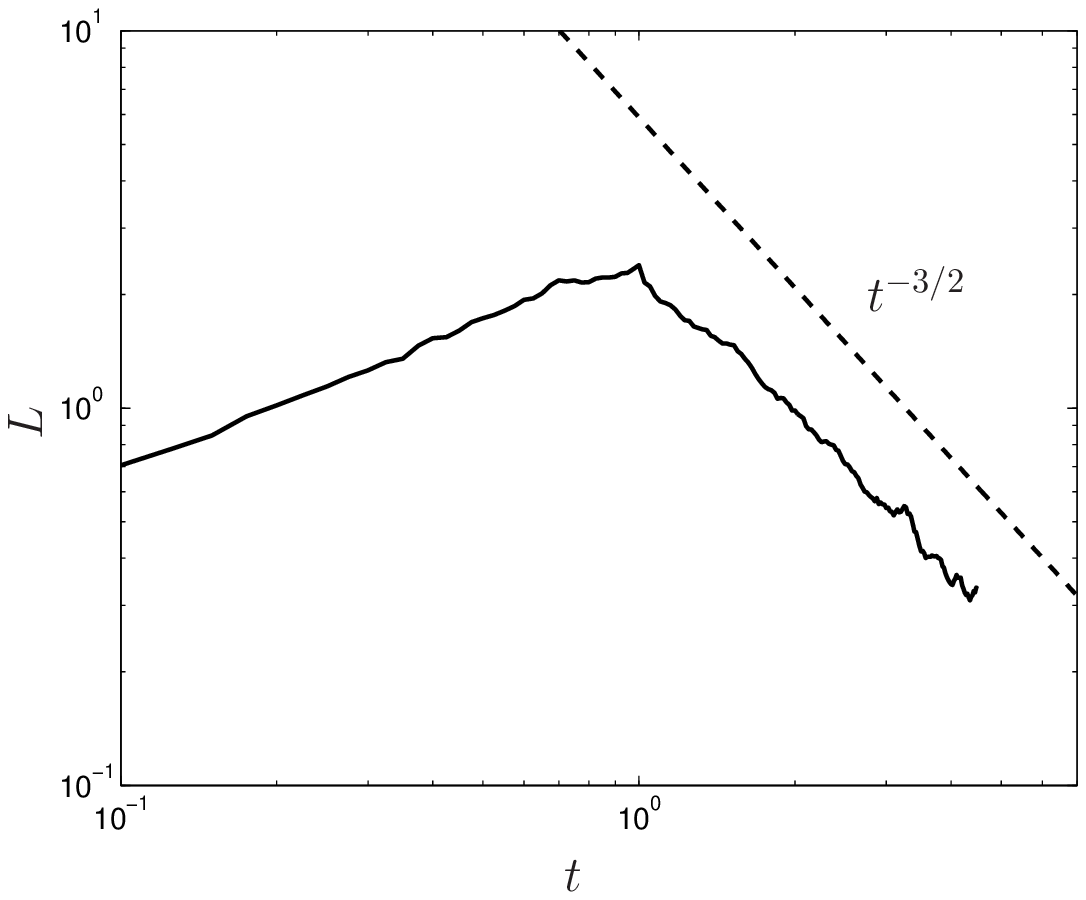}
\caption{
Vortex line density $L$ ($\rm cm^{-2}$) vs time $t$ ($s$)
corresponding to small (left) and large (right) {\bf times} of vortex rings
injection. Fitting the decay as $L\sim t^{-\alpha}$, we obtain
$\alpha=1.07$ and 1.48, respetively. The dashed lines show the 
ultraquantum $L\sim t^{-1}$ (left) and quasiclassical $L\sim t^{-3/2}$ 
(right) behaviours.
}
\label{fig:1}
\end{center}
\end{figure}

By numerically integrating Eq.~(\ref{eq:BS}) we have found that
the vortex rings interact, reconnect and, as envisaged by Bradley
\etal~\cite{Bradley2005}, form a tangle; the vortex line density
$L$ reaches a peak and then decays, see Fig.~\ref{fig:1} (left),
in agreement with the
observed ultraquantum ($L\sim t^{-1}$) behaviour. During the decay,
the kinetic energy (per unit mass), $E$, has the expected
$E \sim t^{-1}$ behaviour.

We have repeated the calculation with longer injection time, up to
$t=1~\rm s$. The peak value of $L$ is thus about 10 times larger
than in the ultraquantum case, as in the
experiment~\cite{Walmsley2008}. We have found that, as shown in
Fig.~\ref{fig:1} (right), after the initial transient the decay
assumes the quasiclassical ($L \sim t^{-3/2}$) form observed in
the experiments~\cite{Walmsley2008,Bradley2006}. We have also
checked that $E \sim t^{-2}$, as expected. The same quasiclassical
and ultraquantum decays are obtained with half the numerical
resolution along the vortex filaments.

It should be emphasized that left and right of Fig.~\ref{fig:1}
do not represent different stages of turbulence but reproduce two
different experiments~\cite{Walmsley2008} resulting, respectively,
in two different regimes of decay: ultraquantum and 
quasiclassical. The key parameter, determining which of the two
regimes will be realized, is the time of injection of vortex
rings.

Assuming the classical expression $dE/dt=-\nu \omega^2$, where
$\omega$ is the vorticity, and the identification $\omega=\kappa
L$, we interpret the results in terms of an effective kinematic
viscosity $\nu$, which we call $\nu_V$ (``Vinen'') and $\nu_K$
(``Kolmogorov'') respectively for the two regimes
\cite{Walmsley2008}. The values of the effective kinematic
viscosities $\nu_V$ and $\nu_K$ have been obtained as in
Ref.~\cite{Walmsley2008} by fitting respectively $L \approx
B/(\nu_V t)$, where $B=(1/(4 \pi))\ln{(\ell/a_0)}$ and $a_0\approx
10^{-8}~\rm cm$ is the vortex core radius, and $L\approx
(3C)^{2/3} \nu_K^{-1/2} \kappa^{-1} k_1^{-1} t^{-3/2}$, where $2
\pi/k_1$ is the large scale and $C=1.5$ is the Kolmogorov
constant. In applying these formulae we have taken into account
the fact that for the calculations presented here in the
ultraquantum case the computational box is not entirely full, and
that in the quasiclassical case the largest length scale is of the
order of $0.06~\rm cm$, as visible in ${\rm PDF}(C)$. We obtain
$\nu_V/\kappa \approx 10^{-1}$ and $\nu_K/\kappa \approx 10^{-3}$,
which compare fairly well with Walmsley \& Golov's $\nu_V/\kappa
\approx 0.08$ to $0.1$ and $\nu_K/\kappa \approx0.002$ to $0.01$.

\begin{figure}[h]
\begin{center}
\includegraphics[width=0.35\textwidth]{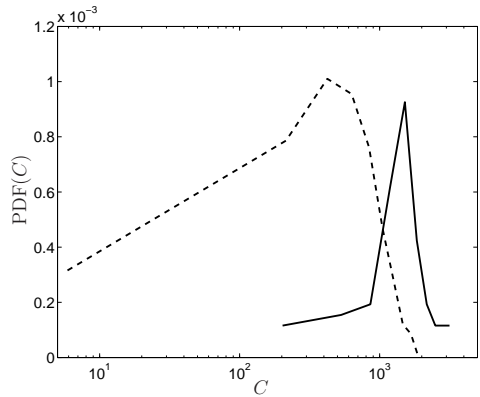}
\caption{Probability distribution functions
of the vortex line curvature,
$C$ ($\rm cm^{-1}$).
Solid line: ultraquantum decay at $t=0.08 ~\rm s$;
dashed line: quasiclassical decay at $t=0.7 ~\rm s$.
}
\label{fig:2}
\end{center}
\end{figure}

To understand the nature of the two regimes we have examined the
time behaviour of the probability density function ${\rm PDF}(C)$
(normalized histogram) of the local vortex line curvature,
$C=\vert d^2 \bs/d\xi^2\vert$. In both ultraquantum and
quasiclassical case, the initial PDF develops in time to larger
and smaller values of $C$. In the quasiclassical case, however,
there is a much greater build up at small values of $C$, see
Fig.~\ref{fig:2}; this means that, as the initial vortex rings
entangle, large-scale structures are created consisting of long
vortex filaments which can extend across the entire computational
domain.

\begin{figure}[h]
\begin{center}
   \includegraphics[width=0.40\textwidth]{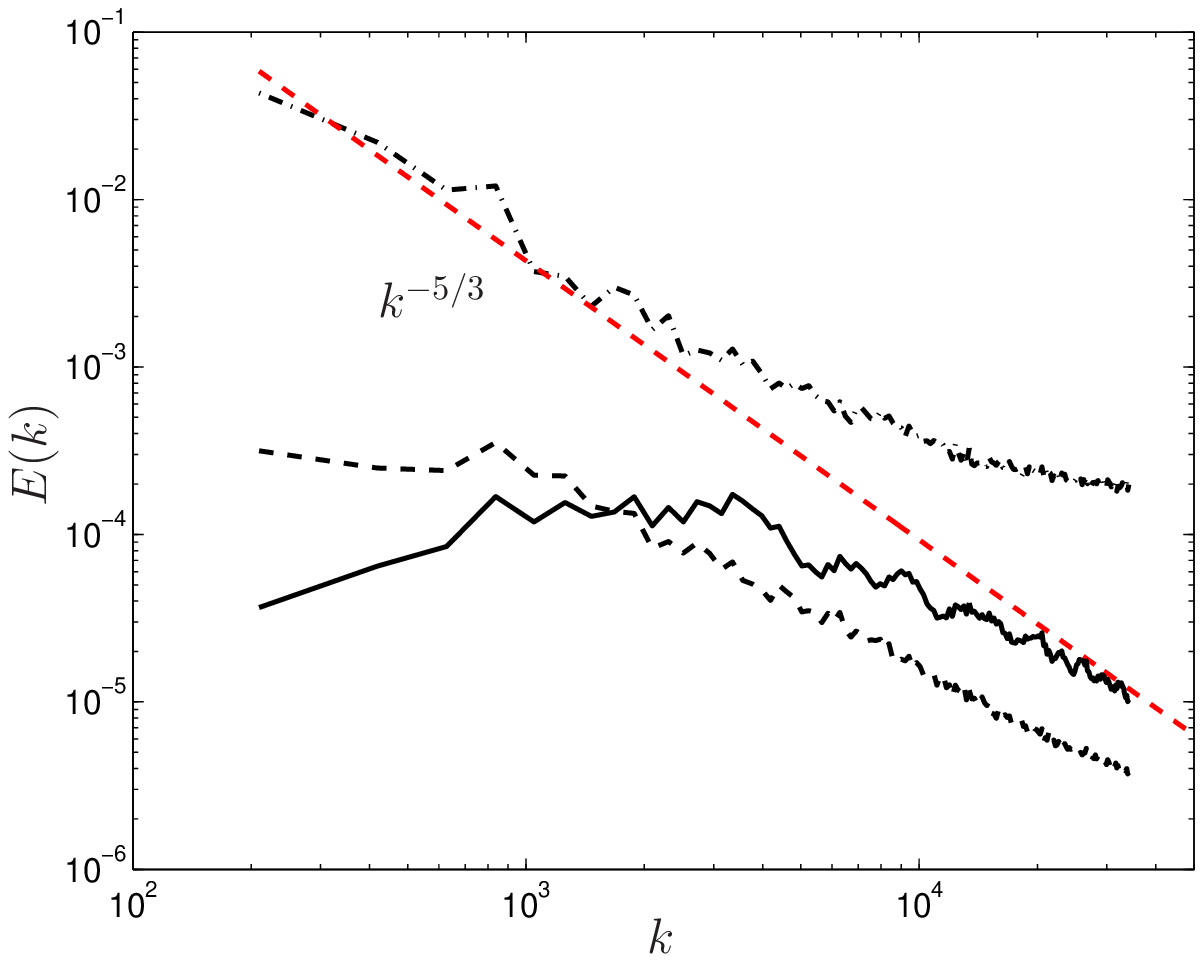}
   \includegraphics[width=0.40\textwidth]{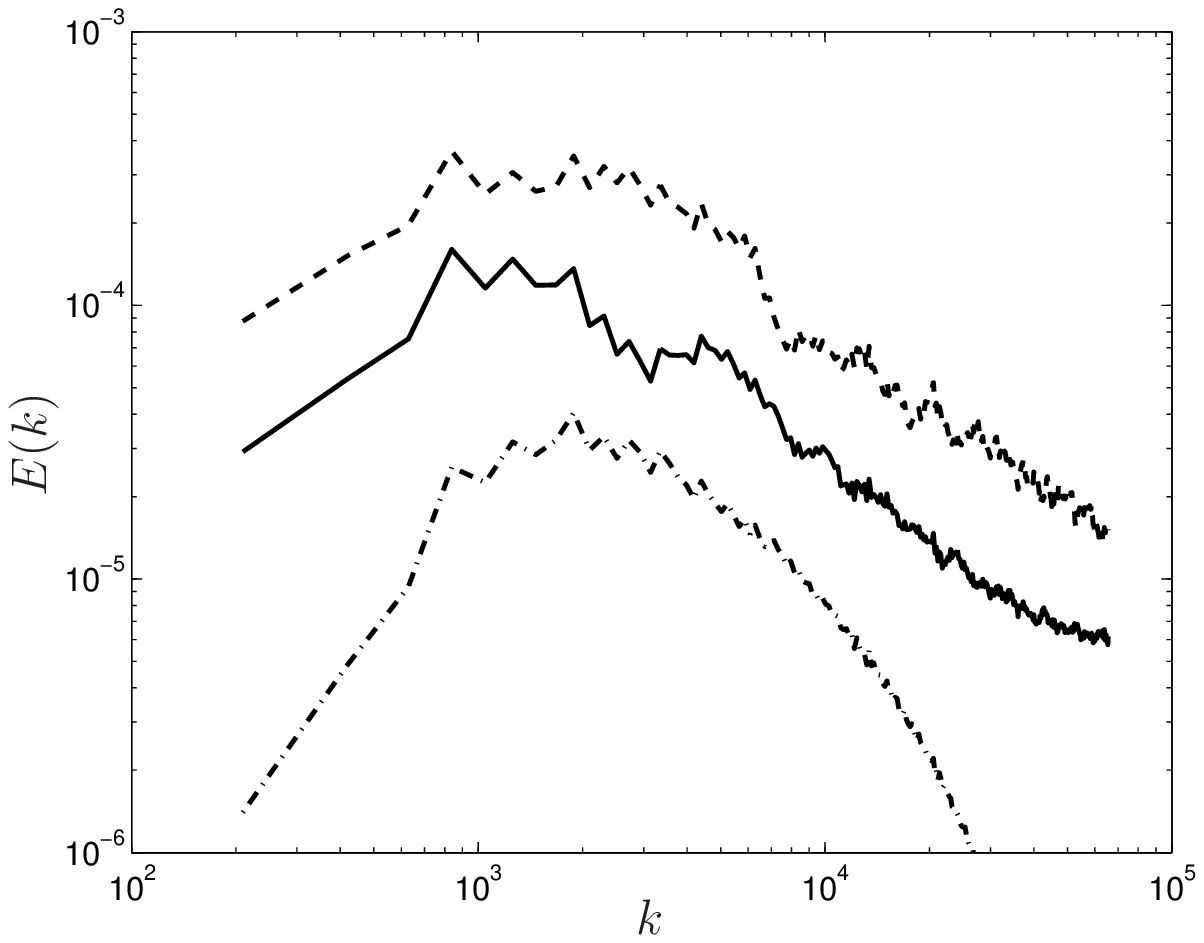}
   \caption{(Color online) Energy spectrum (arbitrary units)
vs wavenumber $k$, $\rm cm^{-1}$ corresponding to the
quasiclassical decay (left) with the solid line at $t=0.1 ~\rm s$,
dot-dashed $t=1.1 ~\rm s$, and dashed $t=3.4 ~\rm s$, and to the
ultraquantum decay (right) with the solid line at $t=0.07 ~\rm s$,
dashed $t=0.12 ~\rm s$, and dot-dashed $t=0.6 ~\rm s$. In the
former, note the formation and decay of the Kolmogorov spectrum
(indicated by the straight dashed line). } \label{fig:3}
\end{center}
\end{figure}

This generation of large length scales is apparent in
Fig.~\ref{fig:3}, where we show the evolution of the kinetic
energy
spectrum $E_k$, defined by
\begin{equation}
E=\frac{1}{V} \int_V \frac{1}{2} \vert\bv\vert^2\, dV=\int_0^{\infty} E_k\, dk
\label{eq:spectrum}
\end{equation}
(where $V$ is volume and $k$ the magnitude of the
three-dimensional wavevector $\bf k$). In both ultraquantum and
quasiclassical cases the energy is initially concentrated at
intermediate wavenumbers. It is apparent (see Fig.~\ref{fig:3}
right) that in the ultraquantum case the value of $k=k_*$, where
$E_k$ has the maximum does not change, and the ratio of energy
transferred to large scales, $\int_0^{k_*}E_k\,dk$, to that
transferred to small scales, $\int_{k_*}^\infty E_k\,dk$, remains
small ($<0.13$) at all times. In the quasiclassical case, however,
a significant amount of energy is transferred to small
wavenumbers, leading to the formation of the Kolmogorov $k^{-5/3}$
spectrum, see Fig.~\ref{fig:3}. The spectrum maintains the
Kolmogorov scaling during the decay stage, consistently with the
observation that $L \sim t^{-3/2}$, even for relatively small
values of $L$ which would otherwise decay as $t^{-1}$ if $L$ were
small initially.

\begin{figure}[h]
\begin{center}
   \includegraphics[width=0.25\textwidth]{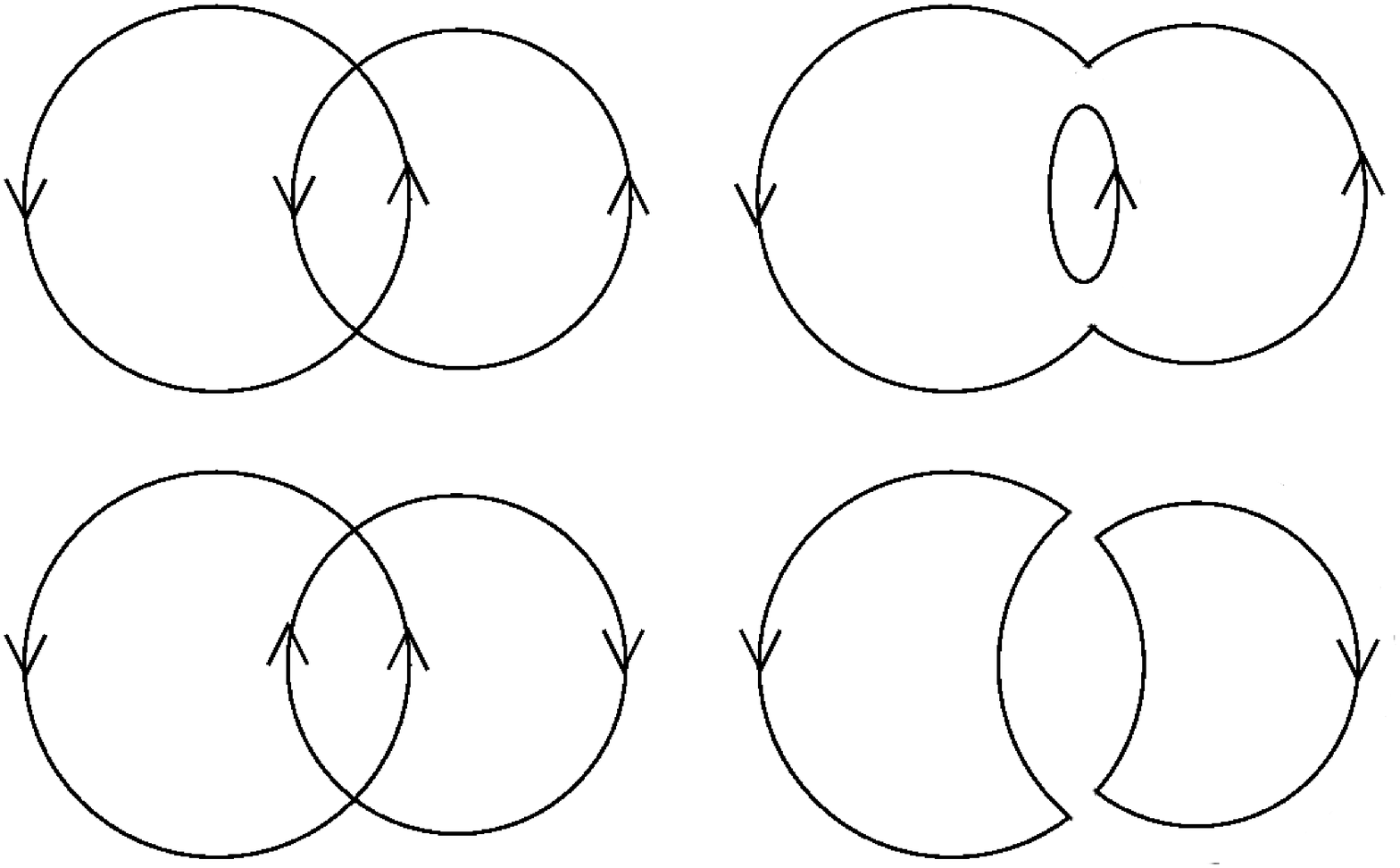}
\caption{
Head-on reconnection of two vortex rings of similar size
tends to produce two vortex loops of similar size (bottom),
whereas the reconnection of two rings traveling in the same
direction produces one vortex loop which is much smaller and one
which is much bigger (top).} \label{fig:4}
\end{center}
\end{figure}
To interpret these results we remark that in both
experiments~\cite{Walmsley2008,Bradley2006} the initial vortex
rings do not move isotropically, but essentially travel in the
same direction as a beam. This anisotropy is important in creating
large length scales, provided that the initial density of the
rings is large enough. The argument is the following. Energy and
speed of a vortex ring of radius $R$ are respectively proportional
and inversely proportional to $R$. Consider the collision of two
vortex rings of approximately the same size. If the collision is
head-on, the outcome of the reconnection will be two vortex loops
of approximately the same size, as shown schematically in
Fig.~\ref{fig:4} (bottom). If the two rings travel
approximately in the same direction, the reconnection will create
two vortex loops, one small and one big, see Fig.~\ref{fig:4} (top).
To test the idea that an anisotropic beam facilitates the
creation of length scales, we have performed numerical
calculations in which the initial distribution of vortex rings
differs only by the orientation of the rings: in one case the
rings pointed isotropically in all directions, and in another case
they pointed in the same direction.
\begin{figure}[h]
\begin{center}
   \includegraphics[width=0.35\textwidth]{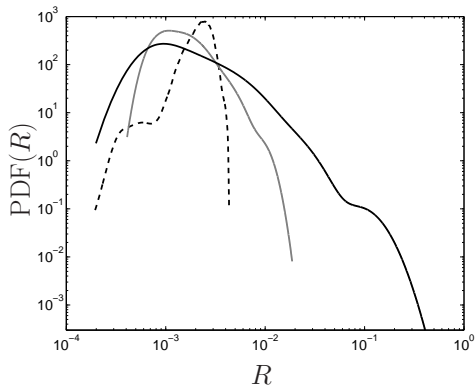}
\caption{ Probability
distribution functions, PDF($R$) of loop sizes $R$, ($\rm cm$).
Dashed line: initial PDF. Grey line: resulting PDF for isotropic
initial condition: no large loops are created by the vortex
reconnections. Black line: resulting PDF for anisotropic initial
condition: note the generation of large loops.} \label{fig:5}
\end{center}
\end{figure} 
Figure~\ref{fig:5} confirms that the anisotropic initial condition generates smaller
values of curvature (that is, larger length scales) as well as
bigger ones. To highlight the geometrical role of vortex
reconnections, the calculation was performed by replacing
Eq.~(\ref{eq:BS}) with the local induction
approximation~\cite{Saffman1992} $d{\bf s}/dt=\beta {\bf s}'
\times {\bf s}''$ (where the prime denotes derivative with respect
to arclength and $\beta$ is constant); in this way the rings
interacted only when they collided. The result was similar to that
obtained using the full Biot-Savart calculation.

In conclusion, our model reproduces both the ultraquantum ($L \sim
t^{-1}$) and the quasiclassical ($L \sim t^{-3/2}$) turbulent
decay regimes which have been observed in the
experiments~\cite{Walmsley2008,Bradley2006} and explains their
hydrodynamical natures. By examining the curvature PDF and the
energy spectrum, we have found that in the quasiclassical regime
the initial energy distribution is shifted to large scales, and a
Kolmogorov spectrum is formed. In the ultraquantum case, the
spectrum decays without this energy transfer. In the case where
turbulence is generated by forcing in the vicinity of some
(intermediate) lengthscale, as in the
experiments~\cite{Walmsley2008,Bradley2006}, we found that the
ultraquantum regime is induced only if the total energy input is
relatively low, while the higher energy input (by e.g. the
prolonged injection of the vortex rings in
experiments~\cite{Walmsley2008,Bradley2006}) generates the large
scale motion and hence the quasiclassical, Kolmogorov regime of
turbulence. We have also found that the anisotropy of the beam of
vortex rings is important, as reconnections of vortex loops
traveling in the same direction are very effective in creating
larger length scales.

\begin{acknowledgments}
We acknowledge the support of
the HPC-EUROPA2 project 228398 (European Community Research Infrastructure
Action of the FP7), the Leverhulme Trust (Grant F/00125/AH), and the
EPSRC (Grant EP/I01941311).
\end{acknowledgments}

\end{document}